\documentclass[twocolumn, switch]{article}
\usepackage{preprint} 
\pdfoutput=1 
\usepackage{authblk}

\usepackage{fancyhdr}
\usepackage{enumitem} \setlist[itemize]{align=parleft,left=0pt..1em}
\usepackage[english]{babel}  
\usepackage{amsmath} 
\usepackage{bm} 
\usepackage{xcolor}
\usepackage{hyperref} 
\usepackage{colortbl, hhline} 
\usepackage{tabularx,booktabs} 
\newcolumntype{Y}{>{\centering\arraybackslash}X} 
\usepackage[superscript,biblabel]{cite}

\definecolor{LightGray}{gray}{.85}

\newcommand{\tI}{\alpha} 
\newcommand{\tII}{\beta} 
\newcommand{\msgood}{\phi} 
\newcommand{\msbad}{\psi} 
\newcommand{\msprofit}{\rho}
\newcommand{\msloss}{\lambda}
\newcommand{\rocop}[1]{o_{#1}}  
\newcommand{\uconf}{k}  
\newcommand{\sens}[1]{\mathit{Se}^{#1}} 
\newcommand{\spec}[1]{\mathit{Sp}^{#1}} 
\newcommand{\TP}{\mathit{TP}} 
\newcommand{\FP}{\mathit{FP}} 
\newcommand{\TN}{\mathit{TN}} 
\newcommand{\FN}{\mathit{FN}} 

\graphicspath{{./}}

\begin{document}

\twocolumn[\begin{@twocolumnfalse}

\title{Statistical validation of a deep learning algorithm for dental anomaly detection in intraoral radiographs using paired data}
\author{Pieter Van Leemput$^*$, Johannes Keustermans and Wouter Mollemans \\ 
Envista -- Medicim N.V., Stationsstraat 102, B-2800 Mechelen, Belgium \\ 
$^*$Corresponding author: \texttt{pieter.vanleemput@envistaco.com}}

\maketitle

\begin{abstract}
This article describes the clinical validation study setup, statistical analysis and results for a deep learning algorithm which detects dental anomalies in intraoral radiographic images, more specifically caries, apical lesions, root canal treatment defects, marginal defects at crown restorations, periodontal bone loss and calculus. The study compares the detection performance of dentists using the deep learning algorithm to the prior performance of these dentists evaluating the images without algorithmic assistance. Calculating the marginal profit and loss of performance from the annotated paired image data allows for a quantification of the hypothesized change in sensitivity and specificity. The statistical significance of these results is extensively proven using both McNemar's test and the binomial hypothesis test. The average sensitivity increases from $60.7\%$ to $85.9\%$, while the average specificity slightly decreases from $94.5\%$ to $92.7\%$. We prove that the increase of the area under the localization ROC curve (AUC) is significant (from $0.60$ to $0.86$ on average), while the average AUC is bounded by the $95\%$ confidence intervals ${[}0.54, 0.65{]}$ and ${[}0.82, 0.90{]}$. When using the deep learning algorithm for diagnostic guidance, the dentist can be $95\%$ confident that the average true population sensitivity is bounded by the range $79.6\%$ to $91.9\%$. The proposed paired data setup and statistical analysis can be used as a blueprint to thoroughly test the effect of a modality change, like a deep learning based detection and/or segmentation, on radiographic images.\\
\end{abstract}

\vspace{0.5cm}
\end{@twocolumnfalse}]

\section{Introduction} \label{S:Intro}

Two-dimensional intraoral radiographs (IOR) are commonly used by dentists for visual inspection of the anatomy of a patient's dentition when a dental disease or issue is suspected. In recent years, deep learning (a.k.a.\ artificial intelligence (AI)) networks have been proposed for the automated detection of dental anomalies in intraoral and panoramic radiographic images, most notably caries \cite{Cantu2020,Khan2021,Lee2018b,Lee2021}, bone loss \cite{Khan2021, Krois2019} and apical lesions \cite{Ekert2019,Celik2023,Hamdan2022}. We have developed a deep learning algorithm to detect and segment the precise location and size of several common dental anomalies visible in IOR images. These anomaly types are caries, apical lesion, a defect at a treated root canal, a marginal defect at a crown restoration, periodontal bone loss and calculus. 

We propose a validation methodology to investigate the effectiveness of the AI-based anomaly detection algorithm for diagnostic assistance in clinical practice. We employ a paired data approach, where each IOR image is evaluated twice by a single dentist, once without (control arm) and once with the help of the AI algorithm (study arm), with a latency period in between both modalities to avoid recall bias. To ensure that the validation results are statistically sound and significant, we formulate hypotheses to prove that the detection sensitivity of the AI-assisted dentists significantly increases, while their detection specificity only slightly decreases. Because of the paired data setup, matched sample tables \cite{Hawass1997} can be calculated to obtain the marginal profit and loss in AI-assisted detection ability. Based on these values, we will show how both McNemar's test \cite{McNemar1947} and binomial probability theory can be used to prove the postulated hypotheses. Because the locations of the anomalies in the IOR image (attributed to various teeth) are important, we will use localization Receiver Operating Characteristic (ROC) curves \cite{He2009} to further visualize the detection performance of the clinicians. Since the Area Under a ROC Curve (AUC) is a widely used performance measure \cite{Fawcett2006}, the hypothesis test and statistics for the AUC from Hanley and McNeil \cite{Hanley1982,Hanley1983} will be used to show that the AUC increase from control to study is large and significant (i.e.\ its $p$-value falls below a preset Type-I error threshold). Finally, we will also derive formulas for the confidence intervals of these statistics.  

\section{Materials} \label{S:Setup}

\subsection{The deep learning network} \label{S:AI}

The AI anomaly detection algorithm uses a convolutional neural network based on the U-Net \cite{Ronneberger2015} architecture, where we replaced the encoding path of the U-Net by a pre-trained VGG19 \cite{Simonyan2015} network. The convolutional neural network produces probabilistic segmentation maps (a.k.a.\ heat map images) for each anomaly, of the same size as the IOR images used as input. These heat maps are converted into binary segmentation masks by thresholding them with anomaly-specific thresholds designed to maximize precision and sensitivity on the tuning data set. Performing connected component analysis with minimal area requirements on these segmentation masks results in the actual detection regions. Typically, these clustered pixel segmentations have an arbitrary shape. For the validation study, tight bounding boxes enclosing the segmented locations were created to represent the AI-generated annotations. 

To tune the parameters of the convolutional neural network, sensitivity was considered twice as important as precision. The data set used for training the deep learning algorithm consists of $3202$ annotated IOR images. An additional set of $936$ IOR images was used for parameter tuning. Both periapical and bitewing IOR images were included. The performance of the AI algorithm itself has been measured on a separate test data set of $452$ IOR images. For both training and comparison, the anomaly locations were manually annotated once by various dental experts. Averaged over the six anomaly categories, the instance-based sensitivity of the AI algorithm itself equals $81.6\%$, with a precision of $46.7\%$ for a tuned $F_2$-score of $70.6\%$.

\begin{figure*}
\centering
\includegraphics[width=0.47\textwidth,keepaspectratio]{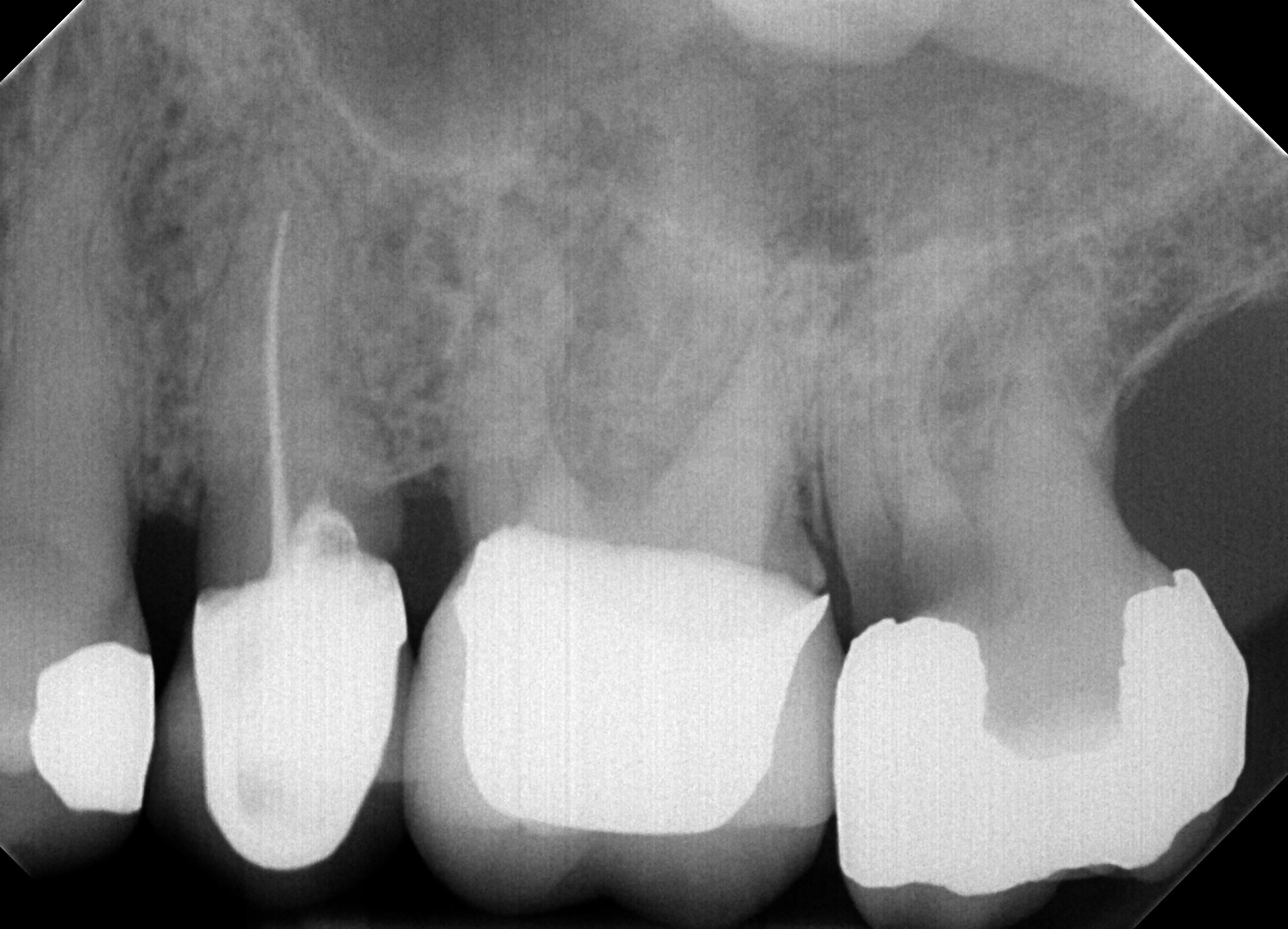}
\hfill
\includegraphics[width=0.47\textwidth,keepaspectratio]{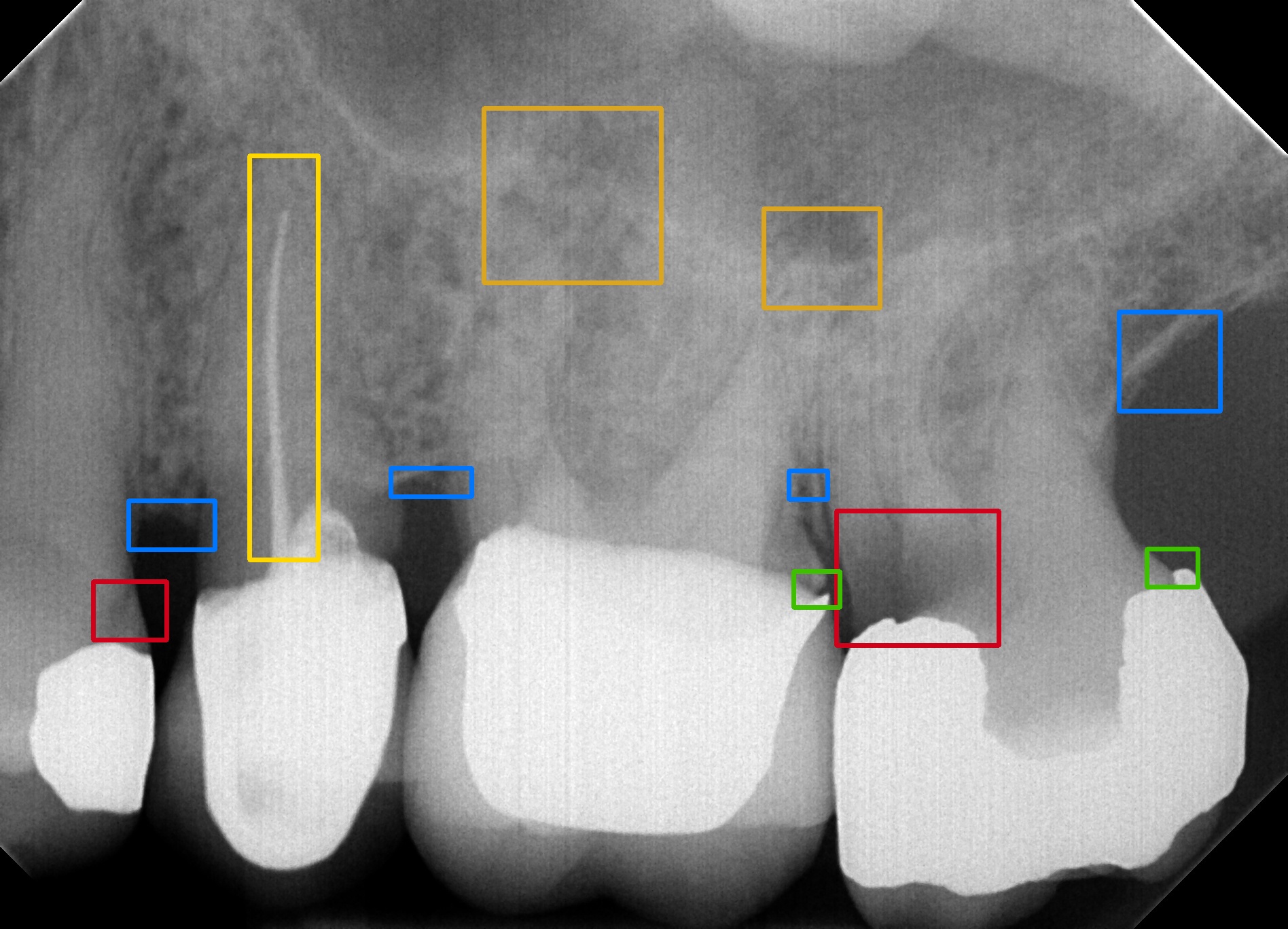}
\caption{Annotation setup for the control arm. The original IOR input image is shown to the left, while its manually annotated counterpart is shown on the right. The color coding of the annotated bounding boxes is as follows: Caries (red), bone loss (dark blue), marginal defect (green), root canal treatment defect (yellow), calculus (light blue) and apical lesion (orange).}
\label{Fig:Annotations}
\end{figure*}

\subsection{Validation data} \label{S:Data}

A representative validation data set of $n_i = 218$ IOR images was selected from an internal company database. These secondary use IOR images were captured in the past by dentists during regular day-to-day dental practice. The IOR images were anonymized and disidentified upon inclusion in the company database, so that it is impossible to link the images (or any anomaly found in the study) back to patients. A total of $182$ IOR images contained instances of at least one of the six anomaly types under investigation. The other $36$ images contained none of these anomalies. The IOR capturing devices were either digital sensors ($186$ images) or photostimulable phosphor (PSP) plates ($32$ images). Molars were depicted about twice as much as the number of incisors and canines combined, with similar distribution numbers for upper and lower teeth.

To evaluate the annotation results from either the control or study arm in \S\ref{S:ControlAndStudy}, the respective annotations must be compared to some form of ground truth about the anomaly locations present in the IOR images. To this end, three dental experts, with either Doctor of Dental Surgery (DDS) or Doctor of Medicine in Dentistry (DMD) qualifications and vast dental experience, each annotated the full set of $n_i$ IOR images following the control arm protocol in \S\ref{S:ControlAndStudy}.  Overlap between these annotations was calculated using the Dice score. In the end, two out of three majority voting was used to decide which anomaly locations were considered to be correct, i.e.\ corresponding to ground truth annotations. 

\subsection{Control and study setup} \label{S:ControlAndStudy}

In the first part of the clinical study, called the control arm, a benchmark was set for the regular workflow of inspecting intraoral images. First of, the $n_i$ IOR images were randomly divided into disjunct subsets of $32$ (or in one case $26$) images. Seven new dentists (with either DDS or DMD qualifications and sufficient clinical practice) were asked to each analyze one of these subsets of unprocessed IOR images for the six anomaly categories listed in \S~\ref{S:Intro}. To facilitate and track the evaluation, these dentists manually annotated the original IOR images by drawing rectangular bounding boxes at the location of the various anomaly instances that they could distinguish. The annotations were made using the online image annotation tool Supervisely (\url{https://supervise.ly}). The bounding boxes for each anomaly type all have a predefined color. An example is shown in Fig.~\ref{Fig:Annotations}.

In the study portion of the clinical validation study, each of the seven dentists was shown the same series of $32$ (or $26$) images as in the control arm. In this case, the images were already analyzed beforehand by the AI-based anomaly detection algorithm. The anomaly locations predicted by the algorithm were also depicted on the IOR image using the setup with colored bounding boxes in the annotation tool. The dentists were asked to look for the six anomaly types in each IOR image in their respective image batch using the algorithmic output for guidance. In practice, the dentists could leave the annotating bounding boxes in place for the anomalies that they were in agreement with. Note that repositioning and resizing the boxes was also possible. Further, they were asked to remove all boxes for anomalies that they judged not to be there. Additionally, they could add boxes for anomalies that they considered to be undetected by the AI algorithm. The end results were stored as the algorithm-assisted annotated images. To avoid any recall bias, an idle period of 4 weeks was implemented in between the control and study phase. Also, we switched the order of the control and study phase for three randomly chosen dentists.

The dentists were also asked to rate all their annotated anomaly instances in both the control and study arm with a confidence score $\uconf$ ranging from $100\%$ to $0\%$ (in increments of $10\%$ for mere practical reasons). These confidence values reflect their belief that what they (or the AI algorithm) annotated on the IOR image is indeed the corresponding anomaly. The $\uconf$ values are categorized as: $100\%$ -- $90\%$ (certainly); $80\%$ -- $70\%$ (most likely); $60\%$ -- $50\%$ (probably); $40\%$ -- $30\%$(potentially); $20\%$ -- $10\%$ (unlikely); $0\%$ (definitely not). It is understood that the dentists only have the IOR image to base their judgment on. Further diagnosis in the patient’s intraoral cavity using standard chairside tooling is not at their disposal for this study. Clearly, the $\uconf = 0\%$ confidence label is only applicable for predictions made by the AI algorithm in the study arm that the dentist does not agree with. In the latter case, the dentists were instead also allowed to remove the AI-based annotation entirely. The final tally is indifferent to either one of these actions. 

\section{Classification} \label{S:Classification}
\subsection{Instance-based classification} \label{S:Classes}

The correspondence between the location, size and anomaly type of the anomaly annotations made in either the control or study arm with respect to the ground truth anomalies was established using the Dice score overlap measure. Afterwards, each anomaly location can be assigned to exactly one of the following classes: i) \emph{false negative}, when the ground truth anomaly location is not annotated by the dentist; ii) \emph{true positive}, when the user-annotated anomaly type and location corresponds to the ground truth type and location or iii) \emph{false positive}, when there is no ground truth anomaly (both location and type) that corresponds to the anomaly annotated by the dentist. 

Because a ground truth is available, a theoretical fourth \emph{true negative} class exists as well. Here, no anomaly is annotated in locations where ground truth anomalies of that type are absent. However, any arbitrary image location without a ground truth annotation could serve as a true negative. As such, the number of true negatives can theoretically be infinite. Note that this is a general issue for object detection tasks.

\subsection{Tooth-based classification} \label{S:ToothLevel}

To remedy the issue with instance-based true negatives, we opted for a validation strategy on tooth level, similar to \cite{Cantu2020} and to some extent \cite{Ekert2019,Logicon,Lee2018b}. To this end, all IOR images were fully partitioned into regions of interest corresponding to the different teeth using a semi-automated segmentation procedure. The total number of segmented teeth over all $n_i$ images equals $n_t = 1346$. Any anomaly location that has been annotated in the IOR image was then assigned to at least one tooth using these segmented regions.

Afterwards, we redefined the previous classification, where classes are assigned in the strict order listed below. Specifically, for each of the six anomaly types, we label each tooth present in the images as either a
\begin{itemize}
\item \emph{False negative} ($\FN$): At least one false negative anomaly instance of that type is allocated to the tooth. Additionally, there could also be true positive or false positive annotations of that type assigned to the tooth. False negative detections are considered the worst because any untreated, undetected anomaly could cause serious discomfort for the patient in the long term.
\item \emph{True positive} ($\TP$): True positive anomaly instances of that type are allocated to the tooth. There could also be false positive annotations present for that anomaly type. Because the true positive detection warrants inspection of the respective tooth, any additional false positive detection on the same tooth is expected to be verified for its validity during treatment of that tooth. 
\item \emph{False positive} ($\FP$): There are no ground truth anomaly instances of the current type assigned to the tooth, but false positive annotations where made by the dentist or AI algorithm. During the next patient visit, the dentist will verify such a false positive finding and most likely not perform the corresponding treatment.
\item \emph{True negative} ($\TN$): No ground truth anomaly instances nor matching clinical annotations of that type are assigned to the tooth.
\end{itemize}

For each of the six anomaly types, the number of true (resp.\ false) positives (resp.\ negatives) counted over all teeth in the images can be summarized in a decision matrix as in Table~\ref{T:DM}. These counts (denoted by $| \cdot |$) are tallied for the control arm and study arm separately. The total teeth count $n_t = |P| + |N|$ can be split in the number of positives $|P|$ where an anomaly is present and the remaining number of negatives $|N|$ for the teeth without anomalies of that type.

\begin{table}
\caption{Decision matrix for a single anomaly type in a validation study with paired data. The notation $|\cdot^c| \rightarrow |\cdot^s|$ represents the counts of true or false negatives ($\TN$, $\FN$) or true or false positives ($\TP$, $\FP$) obtained in the control phase $c$ (dentist only) versus the study phase $s$ (AI-assisted dentist).}
\label{T:DM}
\centering
\begin{tabularx}{0.39\textwidth}{ r | c c }
\toprule
\multicolumn{1}{c|}{Anomaly} & Not present & Present \\ \midrule
Not detected & $|\TN^c| \rightarrow |\TN^s|$ & $|\FN^c| \rightarrow |\FN^s|$ \\ 
Detected & $|\FP^c| \rightarrow |\FP^s|$ & $|\TP^c| \rightarrow |\TP^s|$ \\ 
Total & $|N|$ & $|P|$ \\
\bottomrule
\end{tabularx}
\end{table}

\section{Performance measures} \label{S:Measures}

\subsection{Sensitivity and specificity}

Based on the counts from Table~\ref{T:DM}, the sensitivity $\sens{}$ (a.k.a.\ probability of detection) and specificity $\spec{}$ (a.k.a.\ false negative rate) are typical measures to quantify the investigated performance. They are defined as follows for each anomaly type individually
\begin{equation} \label{sens-spec}
\sens{} = \frac{|\TP|}{|P|} = \frac{|\TP|}{|\TP| + |\FN|} \; \mbox{and} \;
\spec{} = \frac{|\TN|}{|N|} = \frac{|\TN|}{|\TN| + |\FP|} \, .
\end{equation}

To support the claim that the automatic anomaly detection algorithm helps the dentist in evaluating the IOR image, the number of true positives in the study arm should go up while the number of false negatives decreases compared to the control arm. Meanwhile, the number of true negatives and false positives should ideally stay more or less the same. As such, we would like to prove the hypotheses that the sensitivity of the study arm is larger than that of the control arm, while the specificity does not change too much, i.e. 
\begin{equation} \label{clinical_hypothesis}
\sens{s} > \sens{c} \, , \quad \mbox{while} \quad \spec{s} \approx \spec{c} \, .
\end{equation}

\subsection{Matched sample tables} \label{S:MST}

\begin{table}
\caption{Matched sample table for the sensitivity of a single anomaly type (control $c$ vs.\ study $s$). In all cases, a ground truth anomaly is present.}
\label{T:Sens}
\centering
\begin{tabularx}{0.49\textwidth}{ r | c c | c }
\toprule
\multicolumn{1}{c|}{Anomaly} & Not detected ($s$) & Detected ($s$) & Total \\ \midrule
Not detect.\ ($c$) & $|\msbad| \,{=}\, |\FN^s {\cap} \FN^c|$ & $|\msprofit| \,{=}\, |\TP^s {\cap} \FN^c|$ & $|\FN^c|$ \\ 
Detected ($c$) & $|\msloss| \,{=}\, |\FN^s {\cap} \TP^c|$ & $|\msgood| \,{=}\, |\TP^s {\cap} \TP^c|$ & $|\TP^c|$ \\ \hline
Total  & $|\FN^s|$ & $|\TP^s|$ & $|P|$ \\
\bottomrule
\end{tabularx}
\end{table}

\begin{table}
\caption{Matched sample table for the specificity of a single anomaly type (control $c$ vs.\ study $s$). In these cases, a ground truth anomaly is absent.}
\label{T:Spec}
\centering
\begin{tabularx}{0.49\textwidth}{ r | >{\columncolor{LightGray}[0.5\tabcolsep]} c >{\columncolor{LightGray}[0.5\tabcolsep]} c | >{\columncolor{LightGray}[0.5\tabcolsep]} c }
\toprule
\multicolumn{1}{c|}{Anomaly} & \multicolumn{1}{c}{Not detected ($s$)} & \multicolumn{1}{c|}{Detected ($s$)} & \multicolumn{1}{c}{Total} \\ \midrule
Not detect.\ ($c$) & $|\msgood| \,{=}\, |\TN^s {\cap} \TN^c|$ & $|\msloss| \,{=}\, |\FP^s {\cap} \TN^c|$ & $|\TN^c|$ \\
Detected ($c$) & $|\msprofit| \,{=}\, |\TN^s {\cap} \FP^c|$ & $|\msbad| \,{=}\, |\FP^s {\cap} \FP^c|$ & $|\FP^c|$ \\ \hline
Total & $|\TN^s|$ & $|\FP^s|$ & $|N|$ \\
\bottomrule
\end{tabularx}
\end{table}

Matched sample tables \cite{Hawass1997} can be used to derive additional information from a validation study with paired data. Looking at only the positive ground truth anomalies $P$, the matched sample table for sensitivity is constructed as in Table~\ref{T:Sens}. The diagonal counts $|\msgood|$ and $|\msbad|$ mark the unchanged good and bad performance respectively. Of interest here are the anti-diagonal marginal counts $|\msprofit| = |\TP^s \cap \FN^c|$ and $|\msloss| = |\FN^s \cap \TP^c|$. The former represents the improvement (or profit) of the AI-assisted dentist over the unassisted dentist, while the latter tallies the corresponding deterioration (or loss). Ideally, the intersection $\msprofit = \TP^s \cap \FN^c$ is large, while the subset $\msloss = \FN^s \cap \TP^c$ is small.

Similarly, a matched sample Table~\ref{T:Spec} for the ground truth negatives $N$ can be constructed as well. Compared to Table~\ref{T:Sens}, the profit $\msprofit$, loss $\msloss$, good $\msgood$ and bad $\msbad$ subsets are ordered differently. In this case, the improvement is defined by the subset $\msprofit = \TN^s \cap \FP^c$ while the deterioration is measured by the number of elements in $\msloss = \FP^s \cap \TN^c$. Ideally, the anti-diagonal counts $|\msprofit|$ and $|\msloss|$ are roughly equal and small, in line with \eqref{clinical_hypothesis}. 

\subsection{ROC curves} \label{S:ROC}

A Receiver Operating Characteristic (ROC) curve visualizes the performance of a detection task at various operating points $\rocop{\uconf}$ by plotting the sensitivity $\sens{}(\rocop{\uconf})$ with respect to the false positive rate $1 - \spec{}(\rocop{\uconf})$.  Thresholding the confidence scores $\uconf$ set by the dentists (see \S~\ref{S:ControlAndStudy}) at respectively $\uconf = 100\%, \uconf \geq 90\%, \ldots$ up to $\uconf \geq 10\%$ naturally leads to the definition of 10 discrete operating points $\rocop{100}, \rocop{90}, \ldots, \rocop{10}$, constituting the ROC curve. Here, $\sens{}(\rocop{\uconf})$ and $1 - \spec{}(\rocop{\uconf})$ are calculated \eqref{sens-spec} using the partial $|\TP|$, $|\FP|$, $|\TN|$ and $|\FN|$ counts based on only those annotations with a confidence label $k$ above the $\rocop{\uconf}$ threshold. The performance of similar detection tasks, like the control versus the study arm detections, can be evaluated by comparing the corresponding ROC curves. A typical measure for comparison is the Area Under the Curve (AUC) $a$ \cite{Fawcett2006}, which ideally is large. 

For dental treatment, the localization of the individual anomaly instances on specific teeth by the dentist and/or the AI algorithm is important. Typically also, multiple anomaly instances occur in the same IOR image. In line with our tooth-based classification approach, a \emph{localization} ROC analysis (LROC) is most suitable \cite{He2009}. 

\section{Statistical analysis} \label{S:Stats}

\subsection{Hypothesis testing} \label{S:Hypo}

The matched sample tables \ref{T:Sens} and \ref{T:Spec} allow us to quantify the improvement of the clinician using the algorithmic predictions (in the study arm) with respect to the same clinician processing the same IOR image without algorithmic assistance (in the control arm). The hypothesis test \eqref{clinical_hypothesis} for respectively sensitivity and specificity can be reformulated in terms of the marginal profit $|\msprofit|$ and loss $|\msloss|$ counts as follows \cite{Hawass1997}
\begin{equation} \label{marginal_hypothesis}
H_0: |\msprofit| \approx |\msloss| \quad \mbox{and} \quad H_1: |\msprofit| > |\msloss| \; ,
\end{equation}
with $H_0$ the null hypothesis and $H_1$ the corresponding one-sided alternative hypothesis. In line with \eqref{clinical_hypothesis}, we want to prove that the alternative $H_1$ is true for the sensitivity, while we are content with the null hypothesis for specificity. When the specificity would decrease significantly (i.e.\ when $\spec{s} < \spec{c}$) it is more appropriate to test the left-side alternative hypothesis $H'_1: |\msprofit| < |\msloss|$ instead of $H_1$. For simplicity, we will test the one that is conform the observed behavior in the validation data set. The hypotheses \eqref{marginal_hypothesis} can be verified using either the one-sided McNemar's test or the binomial test explained below.  

\subsection{One-sided McNemar's test} \label{S:McNemar}

To test hypothesis \eqref{marginal_hypothesis}, McNemar's chi-squared test statistic $\chi^2$ can be derived \cite{McNemar1947,Hawass1997,Edwards1948} from the marginal counts in our study as 
\begin{equation} \label{mcnemar}
\chi^2 = \frac{\left( | ( |\msprofit| - |\msloss| ) | - 1 \right)^2}{|\msprofit| + |\msloss|} \; ,
\end{equation}
where the outer $|\cdot|$ in the nominator denotes the absolute value. By definition, the test statistic $\chi^2$ \eqref{mcnemar} is two-sided and agnostic to the direction of change. Effectively, McNemar's test verifies if $|\msprofit| \not\approx |\msloss|$, i.e.\ it checks both alternative hypothesis $H_1: |\msprofit| > |\msloss|$ and $H'_1: |\msprofit| < |\msloss|$. 

Next, the probability $\mathcal{P}_{\chi,0}(X \geq \chi^2)$ corresponding to the chi-squared distribution with one degree of freedom is calculated. For a one-sided test as in \eqref{marginal_hypothesis}, the significance $s(\chi^2)$ (a.k.a.\ the $p$-value) of the observation $\chi^2$ \eqref{mcnemar} under $H_0$ is defined as half this value. When $s(\chi^2)$ is smaller than a predefined threshold $\tI = 5\%$ (also called the significance level), i.e. when
\begin{equation} \label{hypoTest_mcnemar}
s(\chi^2) = \frac{1}{2}\; \mathcal{P}_{\chi,0}(X \geq \chi^2) < \tI \; ,
\end{equation}
we reject the null hypothesis $H_0$ from \eqref{marginal_hypothesis} in favor of the one-sided alternative hypothesis $H_1$ (or $H'_1$). When the significance is larger than the threshold $\tI$, there is much support for the null hypothesis and the alternative hypothesis is disproved.  

\subsection{One-sided binomial test} \label{S:BinTest}

The hypothesis \eqref{marginal_hypothesis} can also be tested using binomial probability distributions $\mathcal{B}(n,p)$. First, we reformulate \eqref{marginal_hypothesis} using classical binomial notation to
\begin{equation} \label{marginal_hypothesis_binomial}
H_0: p = p_0 = 0.5 \; , \quad \mbox{and} \quad H_1: p = p_1 > p_0 \; ,
\end{equation}
where $p \in (0,1)$ denotes the probability of success for a single instance. For a given anomaly type, success for sensitivity (resp.\ specificity) corresponds to a true positive (resp.\ true negative) classification of a tooth in the study arm that was labeled as a false negative (resp.\ false positive) before in the control arm. The sample size for the binomial test equals $n = |\msprofit| + |\msloss|$, while the binomial test statistic (i.e.\ the observation) equals either $x = |\msprofit|$ or $x = |\msloss|$ depending on which is the larger of the two values.
This way, the right side alternative hypothesis $H_1$ from \eqref{marginal_hypothesis} is tested when $x = |\msprofit|$, while effectively the left side alternative $H'_1: |\msprofit| < |\msloss|$ is examined when $x = |\msloss|$ is chosen. With the latter choice, the (right-side) statistical analysis for $H_1: p > p_0$ defined in this subsection can simply be reused because the underlying probability distributions are symmetrical.

As in \eqref{hypoTest_mcnemar}, we test if the one-sided binomial significance $s(x)$ (a.k.a.\ $p$-value) is smaller than the predefined significance threshold $\tI$. Mathematically, we have
\begin{equation} \label{hypoTest_binomial}
s(x) = \mathcal{P}_{b,0}(X \geq x) < \tI \; ,
\end{equation}
where $\mathcal{P}_{b,0}$ denotes the discrete binomial probability \cite{Biostatistics} corresponding to the null hypothesis $H_0$. When \eqref{hypoTest_binomial} is true, we reject the null hypothesis $H_0$ in favor of the alternative hypothesis $H_1$  (or $H'_1$). The probability of incorrectly rejecting the null hypothesis is called a Type-I error $e_I$. It follows that $e_I = \tI$. It can be checked experimentally that the $s(\chi^2)$ \eqref{hypoTest_mcnemar} and $s(x)$ \eqref{hypoTest_binomial} values are approximately the same. 

For the one-sided hypothesis test \eqref{marginal_hypothesis_binomial}, the significance threshold $\tI$ corresponds to a probability area under the binomial curve $\mathcal{B}(n,p_0)$ that is delimited on the abscissa by the so-called critical region $[x_{\tI},n]$, where \cite{Biostatistics}
\begin{equation} \label{x_alpha}
x_{\tI} = n p_0 + z_{(1-\tI)} \sqrt{n p_0 (1-p_0)} + 0.5 \; , 
\end{equation}
with $z_{(1-\tI)}$ the $(1-\tI)$ percentage point of the standardized normal distribution $\mathcal{N}(0,1)$, e.g.\ $z_{0.95} = 1.64$. Note that \eqref{x_alpha} should effectively be rounded to a discrete integer value. 

Further, a Type-II error $e_{II}$ is defined as the probability that we do not reject the null hypothesis $H_0$ at a chosen significance level $\tI$, when in reality $H_1$ (or $H'_1$) holds. When the alternative hypothesis is true, it is desirable that, for a given threshold of e.g.\ $\tII = 10\%$: 
\begin{equation} \label{type2}
e_{II} = \mathcal{P}_{b,1}(X < x_{\tI}) < \tII \; ,
\end{equation}
where the probability is evaluated under the alternative binomial distribution $\mathcal{B}(n,p_1)$, with the best estimate for $p_1 = x/n$. Finally, the power of the hypothesis test is defined as the probability to choose the alternative hypothesis when it is in fact true. It is calculated as $100\% - e_{II}$ and should be larger than $100\%-\tII$ to claim with sufficient confidence that $H_1$ (or $H'_1$) effectively holds.

\subsection{Confidence intervals} \label{S:Confidence}

Once a study is performed on a sample of a certain size (here $|P|$ and $|N|$), one obtains values for the sample statistics (here $\sens{}$ and $\spec{}$ \eqref{sens-spec}) which are valid for that particular study. Additionally, it is of interest to make a statement about the value of the unknown statistics $\overline{\sens{}}$ and $\overline{\spec{}}$ that describe the full population. These can be bounded by $100c\%$ confidence intervals $[\sens{} - d_P, \sens{} + d_P]$ and $[\spec{} - d_N, \spec{} + d_N]$ derived from the sample results. Because both $\sens{}$ and $\spec{}$ can be interpreted as binomial success probabilities, we have \cite{Hajian2014}
\begin{equation} \label{conf_sens-spec}
d_P = z_{\frac{1+c}{2}} \sqrt{\frac{\sens{}(1 - \sens{})}{|P|}} \;\; \mbox{and} \;\; d_N = z_{\frac{1+c}{2}} \sqrt{\frac{\spec{}(1 - \spec{})}{|N|}} \; ,
\end{equation}
with $z_{(1+c)/2}$ the $(1+c)/2$ percentage point of the standardized normal distribution $\mathcal{N}(0,1)$, e.g.\ $z_{0.975} = 1.96$ when $c = 0.95$. 

\subsection{AUC statistics} \label{S:AUC}

Due to its close relation to both the Wilcoxon \cite{Wilcoxon1945} statistic and the Mann-Whitney \cite{MannWhitney1947} $U$-statistic, it can be shown \cite{Bamber1975} that the AUC statistic $A$ itself is normally distributed, i.e.\ $A \sim \mathcal{N}(a, \sigma^2(a))$ with a mean equal to the experimentally measured area $a$. The standard deviation $\sigma(a)$ is estimated to be \cite{Hanley1982}
\begin{equation} \label{stderr_auc}
\sigma(a) = \sqrt{\frac{a (1 - a) + (P - 1)(q_1 - a^2) + (N - 1)(q_2 - a^2)}{P N}} \; ,
\end{equation}
with the probabilistic quantities $q_1$ and $q_2$ defined by
\begin{displaymath}
q_1 = \frac{a}{2 - a} \quad \mbox{and} \quad q_2 = \frac{2 a^2}{1 + a} \; .
\end{displaymath}

It follows that the $100c\%$ confidence intervals for the true AUC $\overline{a^c}$ and $\overline{a^s}$ in the control or study arm are defined as
\begin{equation} \label{conf_auc}
a - z_{\frac{1+c}{2}} \sigma(a) \; < \; \overline{a} \; < \; a + z_{\frac{1+c}{2}} \sigma(a) \; .
\end{equation}

In our paired data study, we explicitly used the same IOR images and the same annotating dentists in the control and study arm. This introduces a correlation $r$ between both annotated data sets. As a result, the standard deviation of the AUC difference $(a^s - a^c)$ reads \cite{Hanley1983} 
\begin{equation} \label{stderr_auc_diff}
\sigma(a^s - a^c) = \sqrt{\sigma^2(a^c) + \sigma^2(a^s) - 2 r \sigma(a^c) \sigma(a^s)} \; .
\end{equation}
The correlation $r$ between $a^s$ and $a^c$ can be retrieved from the look-up table in \cite{Hanley1983}. The latter requires knowledge of the correlation $r_P$ between the positive findings $P$ (of control and study) as well as the correlation $r_N$ between the negative findings $N$. These can be calculated as the Kendall rank coefficients using the concordant ($\msgood$ and $\msbad$) and discordant ($\msprofit$ and $\msloss$) subsets from the matched sample tables \ref{T:Sens} and \ref{T:Spec}, s.t. 
\begin{equation} \label{kendall_correlation}
r_P = \frac{|\msgood| + |\msbad| - |\msprofit| - |\msloss|}{P} \;\; \mbox{and} \;\; r_N = \frac{|\msgood| + |\msbad| - |\msprofit| - |\msloss|}{N} \; .
\end{equation}
The $100c\%$ confidence interval for the true AUC delta $\overline{a^s} - \overline{a^c}$ can be obtained from \eqref{conf_auc} by replacing $a$ with $a^s - a^c$ and by using \eqref{stderr_auc_diff}. 

\begin{table*}
\caption{Comparison of the sensitivity $\sens{}$ and specificity $\spec{}$ values (control $\rightarrow$ study) obtained from the tooth-based categorized counts in Table~\ref{T:DM_all} at the $\rocop{50}$ operating point. Also listed are the corresponding $95\%$ confidence intervals for the population sensitivity $\overline{\sens{}}$ and specificity $\overline{\spec{}}$. The mean values averaged over all anomaly types are given in the last row. All values are expressed in $\%$.}
\label{T:ResAll}
\centering
\begin{tabularx}{0.9\textwidth}{ r | c c | c c }
\toprule
Clinical performance & $\sens{c} \rightarrow \sens{s}$ & $95\%$ confidence intervals & $\spec{c} \rightarrow \spec{s}$ & $95\%$ confidence intervals \\ \midrule 
Caries      & $66.0 \rightarrow 84.9$ & $[58.7, 73.4] \rightarrow [79.3, 90.5]$ & $94.6 \rightarrow 93.2$ & $[93.3, 95.9] \rightarrow [91.7, 94.6]$ \\ 
Apical lesion & $70.4 \rightarrow 90.7$ & $[58.2, 82.5] \rightarrow [83.0, 98.5]$ & $98.7 \rightarrow 97.2$ & $[98.1, 99.3] \rightarrow [96.3, 98.1]$ \\ 
Root canal defect & $71.0 \rightarrow 93.5$ & $[55.0, 86.9] \rightarrow [84.9, 100]$ & $99.2 \rightarrow 98.6$ & $[98.7, 99.7] \rightarrow [98.0, 99.3]$ \\ 
Marginal defect & $33.1 \rightarrow 73.0$ & $[25.9, 40.4] \rightarrow [66.2, 79.8]$ & $97.5 \rightarrow 97.0$ & $[96.6, 98.4] \rightarrow [96.0, 97.9]$ \\ 
Bone loss   & $66.1 \rightarrow 91.1$ & $[61.0, 71.1] \rightarrow [88.0, 94.1]$ & $78.3 \rightarrow 71.8$ & $[75.8, 80.9] \rightarrow [69.0, 74.6]$ \\ 
Calculus    & $57.8 \rightarrow 82.3$ & $[49.8, 65.8] \rightarrow [76.1, 88.5]$ & $98.5 \rightarrow 98.3$ & $[97.8, 99.2] \rightarrow [97.6, 99.1]$ \\ 
\midrule 
Average     & $60.7 \rightarrow 85.9$ & $[51.4, 70.0] \rightarrow [79.6, 91.9]$ & $94.5 \rightarrow 92.7$ & $[93.4, 95.5] \rightarrow [91.4, 93.9]$ \\ 
\bottomrule
\end{tabularx}
\end{table*}

\begin{figure*}
\centering
\includegraphics[width=\textwidth,keepaspectratio]{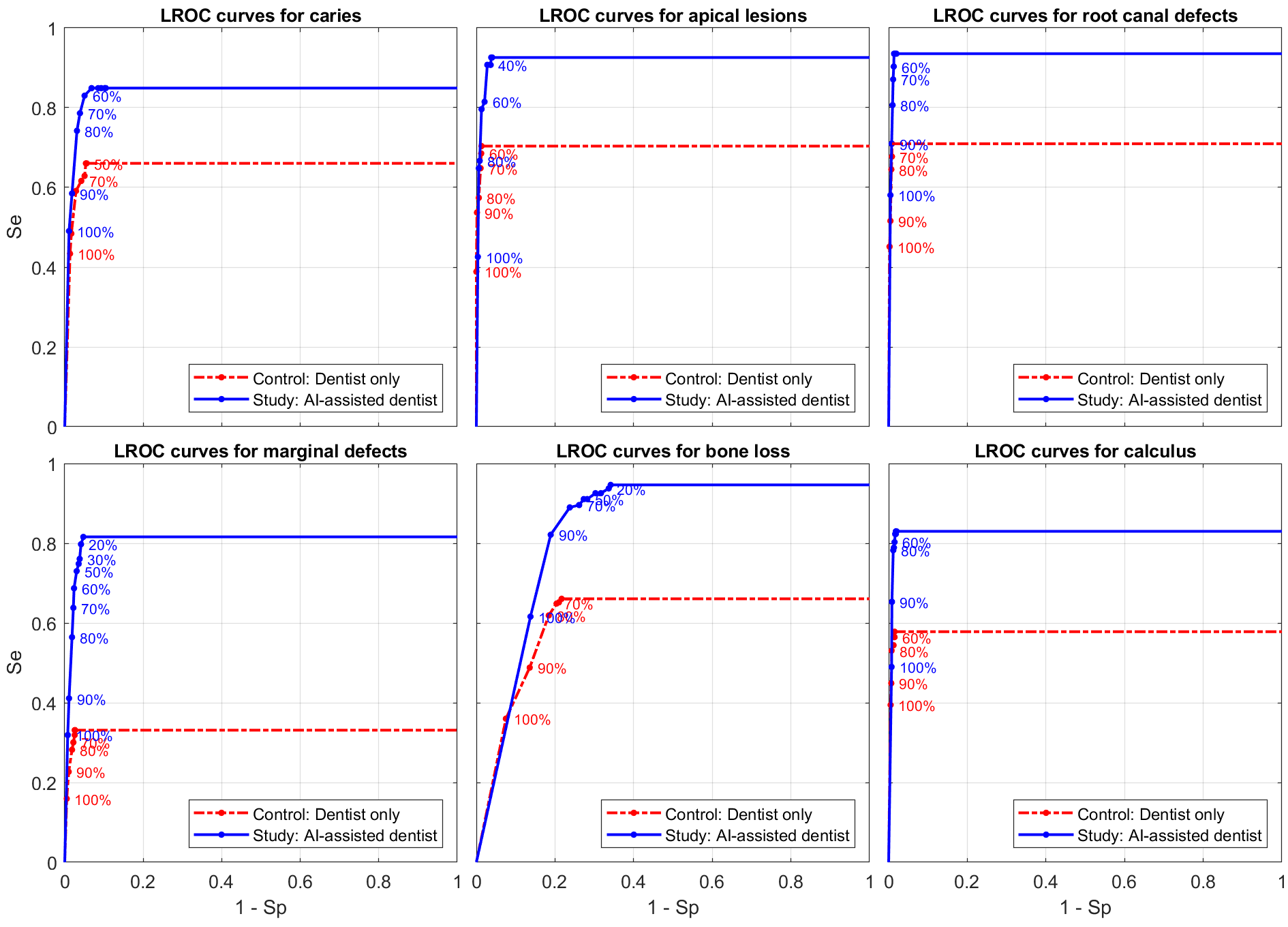}
\caption{The LROC curves derived from the tooth-based validation results in the control and study arm for each of the six anomaly types. A selection of the confidence labels $\uconf$ next to their corresponding discrete operating points $\rocop{\uconf}$ is visualized for clarity.}
\label{Fig:ROC_toothbased}
\end{figure*}

Proving hypotheses \eqref{clinical_hypothesis} implies that the ordinate values $\sens{}$ on a ROC plot increase significantly, while the abscissa values $1 - \spec{}$ stay more or less the same. Because a ROC curve is a monotonically increasing curve which connects the points $\left( 1-\spec{}(\rocop{\uconf}), \sens{}(\rocop{\uconf}) \right)$, this implies that the area under such a curve also increases significantly. Alternatively and more formally,  the following null and one-sided alternative hypotheses can be formulated: 
\begin{equation} \label{hypo_auc}
H_0:  a^s = a^c \quad \mbox{and} \quad H_1: a^s > a^c \; , 
\end{equation}
to test if the change from $a^c$ to $a^s$ is significant. Hanley and McNeil \cite{Hanley1983} propose to test the value 
\begin{equation} \label{z_auc}
\hat{z} = \frac{a^s - a^c}{\sigma(a^s - a^c)} \; ,
\end{equation}
attained by a standardized normally distributed stochastic variable $Z \sim \mathcal{N}(0,1)$. When the one-sided significance (a.k.a.\ $p$-value) $s(\hat{z}) = \mathcal{P}_n(Z \geq \hat{z}) = 1 - \Phi(\hat{z})$ is large, there is much support for the null hypothesis $H_0$ \eqref{hypo_auc}. If $s(\hat{z})$ is small, the alternative hypothesis $H_1$ \eqref{hypo_auc} is true and the change in AUC from control to study is significant. E.g.\ for an allowed Type-I error of $\tI = 5\%$, $H_1$ is deemed true when $\hat{z} > z_{(1 - \tI)} = z_{0.95} = 1.64$. 

\section{Results} \label{S:Results}

All mathematical and statistical results in this text were calculated using MATLAB{\textsuperscript \textregistered} (MathWorks{\textsuperscript \textregistered}). 

Table~\ref{T:ResAll} aggregates the clinical performance results of the validation study. The $\sens{}$ and $\spec{}$ values \eqref{sens-spec} are derived from the decision matrices in Table~\ref{T:DM_all}. Table~\ref{T:ResAll} clearly shows that the sensitivity for the detection of all individual anomalies increases significantly from the control to the study arm, while the specificity values only slightly decrease (except for bone loss). These observations are conform the clinical hypotheses \eqref{clinical_hypothesis} that have been put forward. 

\begin{table}
\caption{Decision matrices listing the tooth-based categorized counts at the $\rocop{50}$ operating point obtained in the control $\rightarrow$ study phases, cf.\ Table \ref{T:DM}.}
\label{T:DM_all}
\centering
\begin{tabularx}{0.41\textwidth}{ r | c c }
\toprule
\qquad\, Caries \qquad\, & Not present & Present \\ \midrule 
Not detected & $1123 \rightarrow 1106$ & $54 \rightarrow 24$ \\ 
Detected     & $64 \rightarrow 81$ & $105 \rightarrow 135$ \\ 
Total & 1187 & 159 \\ 
\bottomrule
\end{tabularx}
\begin{tabularx}{0.41\textwidth}{ r | c c }
\toprule
\;\; Apical lesions \;\; & Not present & Present \\ \midrule 
Not detected & $1275 \rightarrow 1256$ & $16 \rightarrow 5$ \\ 
Detected     & $17 \rightarrow 36$ & $38 \rightarrow 49$ \\ 
Total & 1292 & 54 \\ 
\bottomrule
\end{tabularx}
\begin{tabularx}{0.41\textwidth}{ r | c c }
\toprule
Root canal defects & Not present & Present \\ \midrule 
Not detected & $1304 \rightarrow 1297$ & $9 \rightarrow 2$ \\ 
Detected     & $11 \rightarrow 18$ & $22 \rightarrow 29$ \\ 
Total & 1315 & 31 \\ 
\bottomrule
\end{tabularx}
\begin{tabularx}{0.41\textwidth}{ r | c c }
\toprule 
\,\, Marginal defects & Not present & Present \\ \midrule 
Not detected & $1153 \rightarrow 1147$ & $109 \rightarrow 44$ \\ 
Detected     & $30 \rightarrow 36$ & $54 \rightarrow 119$ \\ 
Total & 1183 & 163 \\ 
\bottomrule
\end{tabularx}
\begin{tabularx}{0.41\textwidth}{ r | c c }
\toprule
\quad\; Bone loss \quad\;\;\, & Not present & Present \\ \midrule 
Not detected & $791 \rightarrow 725$ & $114 \rightarrow 30$ \\ 
Detected     & $219 \rightarrow 285$ & $222 \rightarrow 306$ \\ 
Total & 1010 & 336 \\ 
\bottomrule
\end{tabularx}
\begin{tabularx}{0.41\textwidth}{ r | c c }
\toprule
\quad \;\; Calculus \quad\;\;\; & Not present & Present \\ \midrule 
Not detected & $1181 \rightarrow 1179$ & $62 \rightarrow 26$ \\ 
Detected     & $18 \rightarrow 20$ & $85 \rightarrow 121$ \\ 
Total & 1199 & 147 \\ 
\bottomrule
\end{tabularx}
\end{table}

\begin{table}
\caption{Areas $a^c \rightarrow a^s$ (control vs.\ study) below the LROC curves in Fig.~\ref{Fig:ROC_toothbased}, together with the $95\%$ confidence intervals for the AUC values $\overline{a^c}$ and $\overline{a^s}$ of the population.}
\label{T:AUC_LROC_confInt}
\centering
\begin{tabularx}{0.49\textwidth}{ r | c c }
\toprule
\multicolumn{1}{c|}{AUC statistics} & $a^c \rightarrow a^s$ & $95\%$ confidence intervals \\ \midrule
Caries 					 & $0.65 \rightarrow 0.84$ & $[0.60, 0.70] \rightarrow [0.80, 0.88]$ \\ 
Apical lesion    & $0.70 \rightarrow 0.92$ & $[0.62, 0.78] \rightarrow [0.87, 0.97]$ \\ 
Root canal def.\ & $0.71 \rightarrow 0.93$ & $[0.60, 0.81] \rightarrow [0.87, 0.99]$ \\ 
Marginal def.\   & $0.33 \rightarrow 0.80$ & $[0.29, 0.37] \rightarrow [0.76, 0.85]$ \\ 
Bone loss        & $0.60 \rightarrow 0.84$ & $[0.57, 0.64] \rightarrow [0.81, 0.87]$ \\ 
Calculus         & $0.58 \rightarrow 0.82$ & $[0.53, 0.63] \rightarrow [0.78, 0.87]$ \\
\midrule
Average & $0.60 \rightarrow 0.86$ & $[0.54, 0.65] \rightarrow [0.82, 0.90]$ \\ 
\bottomrule
\end{tabularx}
\caption{The $95\%$ confidence intervals for the $(\overline{a^s} - \overline{a^c})$ difference, together with the hypothesis test values $\hat{z}$ from \eqref{hypo_auc}--\eqref{z_auc} and the $p$-values $s(\hat{z})$.}
\label{T:AUC_LROC_diff}
\centering
\begin{tabularx}{0.44\textwidth}{ r | c c c }
\toprule
\multicolumn{1}{c|}{AUC difference} & $95\%$ conf.~int.\ & $\hat{z}$ & $s(\hat{z})$ \\ \midrule
Caries 						& $[0.15, 0.22]$ & $9.6$ & $0.0$ \\ 
Apical lesion     & $[0.16, 0.28]$ & $7.3$ & $1.4\mbox{e}^{-13}$ \\ 
Root canal defect & $[0.15, 0.30]$ & $6.0$ & $1.1\mbox{e}^{-9}$ \\ 
Marginal defect   & $[0.44, 0.52]$ & $23.1$ & $0.0$ \\ 
Bone loss         & $[0.20, 0.27]$ & $12.9$ & $0.0$ \\ 
Calculus          & $[0.20, 0.30]$ & $10.7$ & $0.0$ \\
\midrule
Average          & $[0.22, 0.31]$ & $-$ & $-$ \\ 
\bottomrule
\end{tabularx}
\end{table}

The $95\%$ confidence intervals \eqref{conf_sens-spec} for control and study sensitivity in Table~\ref{T:ResAll} also exhibit a clear improvement. Because the confidence interval of $\overline{\sens{s}}$ lies to the right of the confidence interval for $\overline{\sens{c}}$ without any overlap (with root canal defects being the inconsequential exception), we can make the following strong statement. With $95\%$ confidence, the true population sensitivity $\overline{\sens{}}$ will increase significantly when the AI algorithm is used to guide the dentist with his/her anomaly detections. 

Figure~\ref{Fig:ROC_toothbased} depicts the LROC curves for each of the six anomaly types. Note that the LROC curve is flattened after the last operating point $\rocop{10}$ due to the localization feature \cite{He2009} and the discrete definition of the operating points. Most LROC curves in Fig.~\ref{Fig:ROC_toothbased} have their operating points clustered together after $\rocop{50}$. In line with the confidence rating scale from \S~\ref{S:ControlAndStudy}, $\uconf \geq 50\%$ marks the decision cutoff for further intraoral dental inspection. Therefore, all counts and performance values in this text are reported at the $\rocop{50}$ operating point. Comparing the control vs.\ study arm values for each pair of LROC operating points $\rocop{\uconf}$ in Fig.~\ref{Fig:ROC_toothbased} clearly shows that the sensitivity $\sens{}$ is high and the false positive rate $1 - \spec{}$ is low (except for bone loss) whenever the AI algorithm is being used. 

The AUC values $a^c$ and $a^s$ that are listed in Table~\ref{T:AUC_LROC_confInt} were calculated using the trapezoidal rule. Tables~\ref{T:AUC_LROC_confInt} and \ref{T:AUC_LROC_diff} also enumerate the AUC confidence intervals as explained in \S~\ref{S:AUC}. In practice, the Kendall rank correlation coefficients \eqref{kendall_correlation} were calculated using the counts from Table~\ref{T:SensAndSpec_all}. The observed AUC increase in Table~\ref{T:AUC_LROC_confInt} (by $26\%$ on average) in the study arm signals a clear advantage when using the AI algorithm for diagnostic assistance. Table~\ref{T:AUC_LROC_diff} shows that the largest $p$-value for the test \eqref{z_auc} equals $s(\hat{z}) = 1.1\mbox{e}^{-9}$. We conclude that in all cases, there is no support for the null hypothesis and the increase $a^s > a^c$ \eqref{hypo_auc} is proven to be significant.

\begin{table}
\caption{Combined matched sample tables with the counts from Table~\ref{T:Sens} (for $\sens{}$) and Table~\ref{T:Spec} (gray cells, for $\spec{}$) at $\rocop{50}$, corresponding to the detection of each of the six anomaly types in the control ($c$) vs.\ study ($s$) phase.}
\label{T:SensAndSpec_all}
\centering
\begin{tabularx}{0.5\textwidth}{ r | Y >{\columncolor{LightGray}} Y Y >{\columncolor{LightGray}} Y | Y >{\columncolor{LightGray}} Y }
\toprule
\multicolumn{1}{c|}{Caries} & \multicolumn{2}{c}{Not detect.\ ($s$)} & \multicolumn{2}{c|}{Detected ($s$)} & \multicolumn{2}{c}{Total} \\ \midrule
Not detect.\ ($c$) & 21 & 1066 & 33 & 57 & 54 & 1123 \\ 
Detected ($c$) & 3 & 40 & 102 & 24 & 105 & 64 \\
\hline
Total & 24 & 1106 & 135 & 81 & 159 & 1187 \\
\bottomrule
\end{tabularx}
\begin{tabularx}{0.5\textwidth}{ r | Y >{\columncolor{LightGray}} Y Y >{\columncolor{LightGray}} Y | Y >{\columncolor{LightGray}} Y }
\toprule
\multicolumn{1}{c|}{Apical lesions} & \multicolumn{2}{c}{Not detect.\ ($s$)} & \multicolumn{2}{c|}{Detected ($s$)} & \multicolumn{2}{c}{Total} \\ \midrule
Not detect.\ ($c$) & 4 & 1247 & 12 & 28 & 16 & 1275 \\
Detected ($c$) & 1 & 9 & 37 & 8 & 38 & 17 \\ \hline
Total & 5 & 1256 & 49 & 36 & 54 & 1292 \\
\bottomrule
\end{tabularx}
\begin{tabularx}{0.5\textwidth}{ r | Y >{\columncolor{LightGray}} Y Y >{\columncolor{LightGray}} Y | Y >{\columncolor{LightGray}} Y }
\toprule
\multicolumn{1}{c|}{Root canal def.} & \multicolumn{2}{c}{Not detect.\ ($s$)} & \multicolumn{2}{c|}{Detected ($s$)} & \multicolumn{2}{c}{Total} \\ \midrule
Not detect.\ ($c$) & 2 & 1295 & 7 & 9 & 9 & 1304 \\
Detected ($c$) & 0 & 2 & 22 & 9 & 22 & 11 \\ \hline
Total & 2 & 1297 & 29 & 18 & 31 & 1315 \\
\bottomrule
\end{tabularx}
\begin{tabularx}{0.5\textwidth}{ r | Y >{\columncolor{LightGray}} Y Y >{\columncolor{LightGray}} Y | Y >{\columncolor{LightGray}} Y }
\toprule
\multicolumn{1}{c|}{Marginal def.} & \multicolumn{2}{c}{Not detect.\ ($s$)} & \multicolumn{2}{c|}{Detected ($s$)} & \multicolumn{2}{c}{Total} \\ \midrule
Not detect.\ ($c$) & 41 & 1125 & 68 & 28 & 109 & 1153 \\
Detected ($c$) & 3 & 22 & 51 & 8 & 54 & 30 \\ \hline
Total & 44 & 1147 & 119 & 36 & 163 & 1183 \\
\bottomrule
\end{tabularx}
\begin{tabularx}{0.5\textwidth}{ r | Y >{\columncolor{LightGray}} Y Y >{\columncolor{LightGray}} Y | Y >{\columncolor{LightGray}} Y }
\toprule
\multicolumn{1}{c|}{Bone loss} & \multicolumn{2}{c}{Not detect.\ ($s$)} & \multicolumn{2}{c|}{Detected ($s$)} & \multicolumn{2}{c}{Total} \\ \midrule
Not detect.\ ($c$) & 20 & 627 & 94 & 164 & 114 & 791 \\ 
Detected ($c$) & 10 & 98 & 212 & 121 & 222 & 219 \\ \hline
Total & 30 & 725 & 306 & 285 & 336 & 1010 \\
\bottomrule
\end{tabularx}
\begin{tabularx}{0.5\textwidth}{ r | Y >{\columncolor{LightGray}} Y Y >{\columncolor{LightGray}} Y | Y >{\columncolor{LightGray}} Y }
\toprule
\multicolumn{1}{c|}{Calculus} & \multicolumn{2}{c}{Not detect.\ ($s$)} & \multicolumn{2}{c|}{Detected ($s$)} & \multicolumn{2}{c}{Total} \\ \midrule
Not detect.\ ($c$) & 13 & 1167 & 49 & 14 & 62 & 1181 \\
Detected ($c$) & 13 & 12 & 72 & 6 & 85 & 18 \\ \hline
Total & 26 & 1179 & 121 & 20 & 147 & 1199 \\
\bottomrule
\end{tabularx}
\end{table}

Table~\ref{T:Stats_sens} summarizes the statistical values for sensitivity as defined in \S~\ref{S:McNemar} and \S~\ref{S:BinTest} for all six anomaly types. The profit $|\msprofit|$ and loss $|\msloss|$ values as defined in Tables~\ref{T:Sens}--\ref{T:Spec} were retrieved from the respective matched sample tables in Table~\ref{T:SensAndSpec_all}. Similarly, Table~\ref{T:Stats_spec} summarizes the statistical validation results for the specificity-related marginal counts $|\msprofit|$ and $|\msloss|$ from Table~\ref{T:SensAndSpec_all} (anti-diagonal gray cells) for all six anomaly types. 

\begin{table}
\caption{Statistical values used to test hypotheses \eqref{marginal_hypothesis} and \eqref{marginal_hypothesis_binomial} for the increase in sensitivity for all anomaly types. The significance values $s(\chi^2)$, $s(x)$, Type-II errors $e_{II}$ and power are expressed as percentages ($\%$).}
\label{T:Stats_sens}
\centering
\begin{tabularx}{0.49\textwidth}{ r | r r | r r r r }
\toprule
\multicolumn{1}{c|}{$\sens{}$ statistics} & $\chi^2$ & $s(\chi^2)$ & $s(x)$ & $x_{\tI}$ & $e_{II}$ & power \\ \midrule
Caries & 23.4 & 0.0 & 0.0 & 23 & 0.0 & 100 \\ 
Lesion & 7.7 & 0.28 & 0.17 & 10 & 1.4 & 98.6 \\ 
Root can.~def.\ & 5.1 & 1.17 & 0.78 & 6 & 0.0 & 100 \\
Margin.~def.\ & 57.7 & 0.0 & 0.0 & 43 & 0.0 & 100 \\
Bone loss & 66.2 & 0.0 & 0.0 & 61 & 0.0 & 100 \\
Calculus & 19.8 & 0.0 & 0.0 & 38 & 0.0 & 100 \\ 
\bottomrule
\end{tabularx}
\end{table}

\begin{table}
\caption{Statistical values used to test hypotheses \eqref{marginal_hypothesis} and \eqref{marginal_hypothesis_binomial} for unchanged specificity for all anomaly types. The significance values $s(\chi^2)$, $s(x)$, Type-II errors $e_{II}$ and power are expressed as percentages ($\%$).}
\label{T:Stats_spec}
\centering
\begin{tabularx}{0.49\textwidth}{ r | r r | r r r r }
\toprule
\multicolumn{1}{c|}{$\spec{}$ statistics} & $\chi^2$ & $s(\chi^2)$ & $s(x)$ & $x_{\tI}$ & $e_{II}$ & power \\ \midrule
Caries & 2.6 & 5.21 & 5.19 & 57 & 45.7 & 54.3 \\ 
Lesion & 8.8 & 0.15 & 0.13 & 24 & 4.7 & 95.3 \\ 
Root can.~def.\ & 3.3 & 3.52 & 3.27 & 9 & 32.2 & 67.8 \\
Margin.~def.\ & 0.5 & 23.98 & 23.99 & 31 & 76.1 & 23.9 \\
Bone loss & 16.1 & 0.003 & 0.003 & 145 & 0.7 & 99.3 \\
Calculus & 0.04 & 42.23 & 42.25 & 18 & 91.7 & 8.3 \\ 
\bottomrule
\end{tabularx}
\end{table}

The sensitivity improvement $\sens{s} > \sens{c}$ observed in Table~\ref{T:ResAll} is significant when the statistical tests \eqref{mcnemar}--\eqref{hypoTest_mcnemar} or \eqref{hypoTest_binomial}--\eqref{type2} support the alternative hypotheses \eqref{marginal_hypothesis} and \eqref{marginal_hypothesis_binomial}. Table~\ref{T:Stats_sens} shows that in all cases, both the chi-squared significance $s(\chi^2)$ and the binomial significance $s(x=|\msprofit|)$ are substantially smaller than the Type-I error threshold $\tI = 5\%$. With the largest value being $s(x)=0.78\%$, we can state that all $p$-values lie well below $0.008$. Additionally, the probability of a Type-II error \eqref{type2} given that $H_1$ holds is always $e_{II} \ll \tII = 10\%$. As a result, the hypothesis test for sensitivity also has more than sufficient power (above $90\%$). All of this implies that we can confidently reject the null hypothesis in favor of the alternative hypothesis, i.e.\ that the increase in sensitivity profit $|\msprofit|$ over loss $|\msloss|$ is real. 

Table~\ref{T:ResAll} reveals a small decrease in the specificity values for most anomalies. To see if this falls within the limits of the null hypothesis $\spec{s} \approx \spec{c}$, we again calculated the statistical tests \eqref{mcnemar}--\eqref{hypoTest_mcnemar} or \eqref{hypoTest_binomial}--\eqref{type2} to obtain Table~\ref{T:Stats_spec}. In line with the observed specificity decrease, the marginal specificity improvement $|\msprofit|$ (not detecting an anomaly which is not there) is smaller than the marginal loss $|\msloss|$ for all anomalies. The latter simply implies that there are more false positive detections in the study arm. As explained in \S~\ref{S:Hypo}, we will actually test the left side alternative hypothesis $H'_1: |\msprofit| < |\msloss|$ in \eqref{marginal_hypothesis} and equivalently $H'_1: p = p_1 < p_0$ in \eqref{marginal_hypothesis_binomial}, by evaluating $s(x = |\msloss|)$ and $p_1 = |\msloss|/n$ for the binomial test statistics \eqref{hypoTest_binomial}--\eqref{type2}. For the detection of caries, marginal defects and calculus, the significance values $s(\chi^2) \approx s(x) > \tI = 5\%$ are large. This implies that there is not enough support for the alternative hypothesis $H'_1: \spec{s} < \spec{c}$ and the null hypothesis $H_0: \spec{s} \approx \spec{c}$ ought not to be rejected. Recall from \S~\ref{S:Hypo} that this is exactly what we wanted to prove for the specificity. Also in these cases, the probability of a Type-II error $e_{II} \gg \tII = 10\%$ is large, which means that when you choose to believe the alternative hypothesis, you are likely making a mistake. Consequently, the power of the specificity test for the aforementioned anomaly types falls way below $90\%$. Note that for the detection of lesions, bone loss and to a lesser extent root canal defects, the statistical values in Table~\ref{T:Stats_spec} show that the specificity decrease is significant, thereby disproving the corresponding null hypothesis. Except for bone loss, the observed decrease in specificity is Table~\ref{T:ResAll} is however small. An increased number of false positives is to be expected when using the AI algorithm, especially since the latter was fine-tuned to avoid false negatives instead of false positives using the $F_2$-score. 

\section{Discussion} \label{S:Discussion}

The quantitative results for the sensitivity, specificity and AUC performance metrics, being supported by detailed statistics, show that the deep learning algorithm from \S~\ref{S:AI} is very capable of providing diagnostic assistance for the detection and localization of the six dental anomalies in IOR images.  

Deep learning networks for dental anomaly detection in IOR and panoramic radiographic images have been previously proposed for the detection of caries \cite{Cantu2020,Khan2021,Lee2018b,Lee2021}, bone loss \cite{Khan2021, Krois2019} and apical lesions \cite{Ekert2019,Celik2023,Hamdan2022}. For the most part, these studies compare the stand-alone AI detection performance directly to the anomaly detections made by several dentists. Lee \emph{et al.} \cite{Lee2021} and Hamdan \emph{et al.} \cite{Hamdan2022} did report on the performance of dental practitioners using their AI algorithm as a diagnostic aid, albeit that Lee \emph{et al.} \cite{Lee2021} has a clear observer bias in their study setup. A more ad hoc approach for caries detection in IOR images was developed by Gakenheimer \cite{Logicon}. Statistical analysis of the validation results reported in these studies is typically limited, leaving some ambiguity in the interpretation of the reported performance measures, a.o.\ by averaging the statistics over various data subsets \cite{Khan2021,Krois2019,Ekert2019, Logicon} and/or observers \cite{Cantu2020, Krois2019,Logicon}; by a lack of details and understanding about classification at instance, tooth or image level; the anomaly localization feature (e.g.\ how to classify two caries instances on one tooth); the ROC curve type or the definition of the ROC operating points \cite{Lee2018b,Krois2019,Ekert2019,Celik2023}.

We have set up a paired data study where the dentists evaluated separate sets of IOR images for the presence of these anomalies. In the control arm, they investigated their subset of images on their own. In the study arm, these dentists evaluated their subset of IOR images a second time, together with the anomaly locations predicted by the algorithm. Because the image and reader set is split up in disjunct parts in our paired data setup, correlations in the validation data only occur unilaterally between the paired control and study arm modalities. This allows for an unbiased aggregation of the validation data over the image and reader ``dimensions'' in each modality, see also \cite{Hanley1983, Metz1989}. In contrast, a fully crossed multireader multicase (MRMC) study design requires all clinical readers to evaluate all images in both the control and study arm modalities. This introduces additional correlations in the validation data which are counteracted by fairly complex and approximate statistical models \cite{Dorfman1992, Hillis2005} employing jackknifing and variance estimates for the unknown population distributions of both readers and images. 

The proposed validation setup and statistical analysis provides an alternative to a MRMC study with some distinct advantages. First, while the number of images can be the same, the paired data approach requires only a fraction of the annotation work of the readers because each (control, study) pair of images is reviewed by only one reader, instead of all readers having to review all images. Second, more traditional statistical analysis \cite{Hanley1982,Hanley1983,Biostatistics} can be used because the reader-evaluated images are treated as independent entities in the paired data setup. A MRMC study on the other hand requires much more complicated statistics that relies on estimates of the variance in the cross-correlated data \cite{Dorfman1992, Hillis2005}. Third, the performance measures and statistics in a paired data study are arguably more relatable because they are directly based on the actual marginal study improvement and loss \cite{Hawass1997} instead of being averaged over a large sample of cross-correlated data in the MRMC model as in e.g.\ \cite{Hamdan2022}. We believe that our study setup and statistical analysis will also be useful to others to test the performance of an AI-based detection or segmentation (or a more general modality change) on radiographic images.

\section{Conclusion}

In this paper, we proposed a validation approach to evaluate the performance of a deep learning algorithm for dental anomaly detection in intraoral radiographic images. The algorithm is able to detect the location and shape of the following six dental anomaly types: caries, apical lesions, root canal treatment defects, marginal defects at crown restorations, periodontal bone loss and calculus. The average sensitivity when using the deep learning algorithm for diagnostic assistance increased significantly from $60.7\%$ to $85.9\%$, while the average specificity only slightly decreased from $94.5\%$ to $92.7\%$. On average, the $95\%$ confidence interval for the true (unknown) population sensitivity improved drastically from $[51.4\%, 70.0\%]$ to $[79.6\%, 91.9\%]$. Using detailed statistical analysis of the performance measures calculated from the marginal profit and loss of the paired data, we have extensively proven the hypotheses that a) the sensitivity always increases significantly and b) the specificity stays more or less the same (except for bone loss) using both McNemar's test and binomial hypothesis testing. Also, the average AUC, i.e.\ the area under the localization ROC curve, increases from $0.60$ to $0.86$, which we proved to be significant. The corresponding average true AUC is bounded by the $95\%$ confidence intervals $[0.54, 0.65]$ and $[0.82, 0.90]$. 

\section*{Acknowledgements and competing interests}
The authors are grateful to the dental experts who participated in the validation study and to Mansour Nadjmi for helping out with annotating and reviewing the AI training data.
Also, the authors are employees of Envista -- Medicim N.V., which released the deep learning algorithm for anomaly detection in the DTX Studio\textsuperscript{TM} Clinic and DEXIS\textsuperscript{TM} 10 Imaging software.

\section*{Author contributions}
\textbf{P.~Van Leemput:} Conceptualization, Methodology, Software, Validation, Formal analysis, Investigation, Writing - original draft, Writing - review \& editing, Visualization, Supervision and Project administration. \textbf{J.~Keustermans:} Methodology, Software, Validation, Investigation, Resources and Data curation. \textbf{W.~Mollemans:} Conceptualization, Project administration and Supervision.  

\bibliographystyle{plain}
\bibliography{aiAnomalyValidationIOR_refs}

\end{document}